\documentclass[aps,prl,showpacs,twocolumn,superscriptaddress,floatfix]{revtex4}

\usepackage{graphicx}
\usepackage{dcolumn}
\usepackage{bm}

\begin{document}

\renewcommand{\vec}[1]{\mbox{\boldmath $#1$}}


\title{Spin-orbit coupling inactivity of Co$^{2+}$ ion in geometrically frustrated magnet GeCo$_2$O$_4$}


\author{Keisuke Tomiyasu}
\email[Electronic address: ]{tomiyasu@m.tohoku.ac.jp}
\affiliation{Department of Physics, Tohoku University, Aoba, Sendai 980-8578, Japan}
\author{Ayaka Tominaga}
\affiliation{Department of Physics, Chuo University, Bunkyo, Tokyo 101-8324, Japan}
\author{Shigeo Hara}
\affiliation{Department of Physics, Chuo University, Bunkyo, Tokyo 101-8324, Japan}
\author{Hirohiko Sato}
\affiliation{Department of Physics, Chuo University, Bunkyo, Tokyo 101-8324, Japan}
\author{Tadataka Watanabe}
\affiliation{Department of Physics, CST, Nihon University, 
Chiyoda, Tokyo 101-8308, Japan}
\author{Shin-Ichi Ikeda}
\affiliation{Nanoelectronics Research Institute, National Institute of AIST, 
Tsukuba 305-8568, Japan}
\author{Kazuaki Iwasa}
\affiliation{Department of Physics, Tohoku University, Aoba, Sendai 980-8578, Japan}
\author{Kazuyoshi Yamada}
\affiliation{WPI AIMR, Tohoku University, 
Aoba, Sendai 980-8577, Japan}


\date{\today}

\begin{abstract}
We report single-crystal neutron diffraction studies on a spinel antiferromagnet GeCo$_2$O$_4$, which exhibits magnetic order with a trigonal propagation vector and tetragonal lattice expansion ($c/a\simeq1.001$) below $T_{\rm N}=21$ K. For this inconsistency between spin and lattice in symmetry, magnetic Bragg reflections with a tetragonal propagation vector were discovered below $T_{\rm N}$. We discuss spin and orbital states of Co$^{2+}$ ion underlying the new magnetic component. 
\end{abstract}


\maketitle

%
%
Geometrical frustration causes mysterious spin-orbital-lattice states. In spinel-type oxides $A$$B_2$O$_4$, where $A$ and $B$ are metal ions, the $B$ site octahedrally surrounded by anions forms a network of corner-sharing tetrahedra, also known as the pyrochlore lattice, which is geometrically frustrated. Recently, exotic phenomena in spinel-type magnets with orbital degree of freedom have attracted much attention. For example, $A$V$_2$O$_4$ ($A$ = Zn, Mn, Li, Al) with $t_{2g}$ orbital degree of freedom exhibit spin and orbital chain, orbital ordering, heavy fermion behavior and heptamer spin singlet state~\cite{Lee_2004,Suzuki_2007,Matsushita_2005,Horibe_2006}.

A spinel-type antiferromagnet GeCo$_2$O$_4$ (Co$^{2+}$: $d^7$, $S=3/2$) is also expected to have degree of freedom in $t_{2g}$ orbital states~\cite{Watanabe_2008}. The cobaltite simultaneously exhibits magnetic long-range order described by a trigonal propagation vector $\vec{Q}_{\rm II}=(1/2,1/2,1/2)$ and cubic-tetragonal lattice expansion at $T_{\rm N}=21$ K ($c/a\simeq1.001$)~\cite{Hubsch_1987,Hoshi_2007}. The spin and lattice symmetries contradict each other. 

A NaCl-type antiferromagnet CoO (Co$^{2+}$: $d^7$, $S=3/2$) once had the same problem: this material also exhibits magnetic order with $\vec{Q}_{\rm II}$ and tetragonal lattice contraction below $T_{\rm N}=290$ K ($c/a\simeq1.01$)~\cite{Shull_1951,Tombs_1950,Germann_1974}. However, careful neutron diffraction experiments later revealed the existence of additional antiferromagnetic order with a tetragonal propagation vector $\vec{Q}_{\rm I}$=(0,0,1)~\cite{Tomiyasu_2004}. Therefore, GeCo$_2$O$_4$ likely hides an additional magnetic order with tetragonal symmetry as well. 

In this paper, we report our discovery of magnetic Bragg reflections with a tetragonal propagation vector by single-crystal neutron diffraction on GeCo$_2$O$_4$. The spin and orbital state of Co$^{2+}$ ion is discussed in comparison with that in CoO. 

%
Neutron diffraction (elastic scattering) experiments were carried out on a triple-axis spectrometer, AKANE (T1-2) of Institute of Materials Research, Tohoku University, installed at the thermal guide tube of JRR-3 in Japan Atomic Energy Agency (JAEA)~\cite{Hiraka_2007} and TOPAN (6G) of the same University, installed at the same reactor. AKANE equips a Ge 311 monochromator and a PG 002 analyzer, and effectively eliminates higher-order contamination without a filter because of a Ge 622 forbidden reflection. Incident energy was fixed at $E_{\rm i}=19.4$ meV with horizontal collimation sequence of guide-open-60$^\prime$-60$^{\prime}$. On TOPAN, $E_{\rm i}$ was fixed at 13.5 meV and 30.5 meV with that of blank-60$^\prime$-60$^{\prime}$-blank. Usually one pyrolytic graphite filter is used to eliminate the half-lambda contamination, but we used two filters to completely eliminate it down to the order of 10$^{-6}$. A single-crystal rod of GeCo$_2$O$_4$ was enclosed in an aluminum container with $^4$He exchange gas, which was mounted on the cold finger of a $^4$He closed-cycle refrigerator. 
The crystal rod was grown by a floating zone method. Detail of the crystal growth was summarized in Ref.~\cite{Hara_2005}. The size of rod is about 4 mm diameter and 20 mm height. 

Since our crystal has multiple tetragonal domains and the $|1-c/a|$ value is only 0.1{\%}, we cannot distinguish 001 and 100 (parallel and perpendicular to the tetragonal axis) in the present experiments. Hereafter we use the cubic notation for arbitrary $hkl$ indices and propagation vectors. 

%

\begin{figure*}[htbp]
\begin{center}
\includegraphics[width=0.95\linewidth, keepaspectratio]{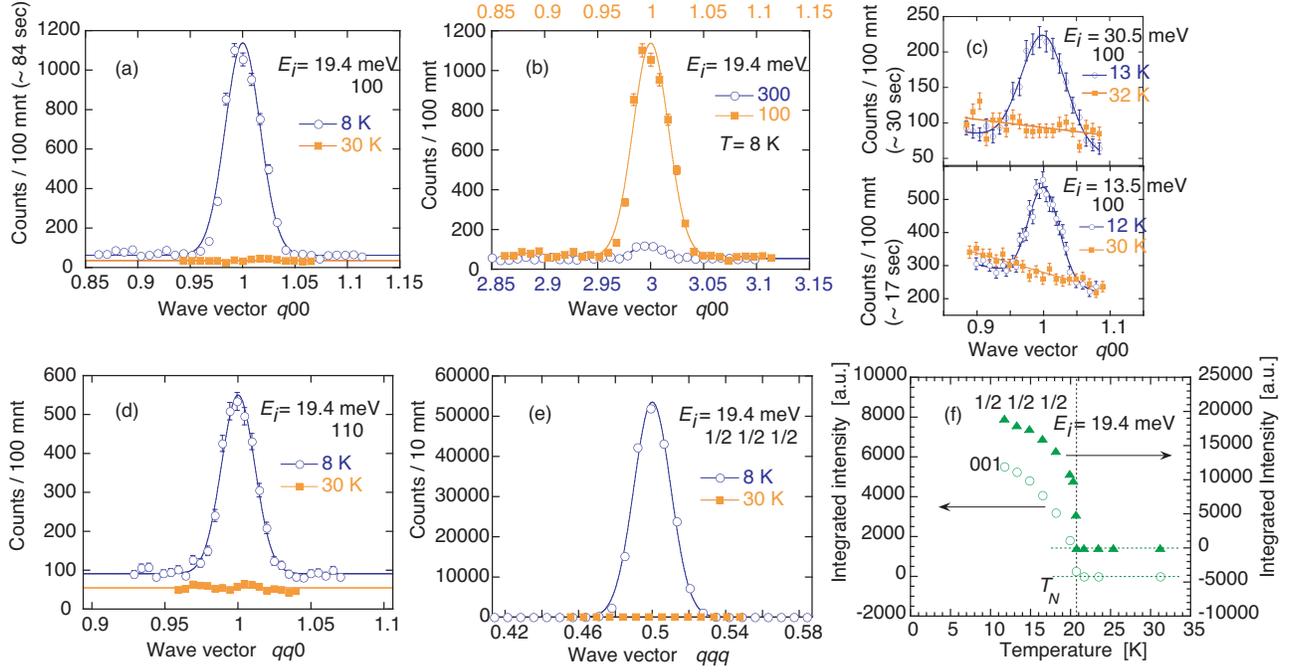}
\end{center}
\caption{\label{fig:data} (Color online) Neutron diffraction data, measured on a single-crystal specimen of GeCo$_2$O$_4$. The 100 reflection (a), comparison between 100 and 300 reflections (b), 100 reflections with different $E_{\rm i}$ (c), 110 reflection (d), 1/2 1/2 1/2 reflection (e) and temperature dependence of integrated intensity of 100 and 1/2 1/2 1/2 reflections (f). 
}
\end{figure*}

Figure~\ref{fig:data}(a) shows the neutron diffraction data, obtained by scanning along the $\langle100\rangle$ direction below and above $T_{\rm N}$ with $E_{\rm i}=19.4$ meV. The unreported reflection is observed at the 100 reciprocal lattice point only below $T_{\rm N}$. Figure~\ref{fig:data}(b) shows the data of 100 reflection and equivalent 300 one below $T_{\rm N}$. The former intensity is much stronger than the latter one, indicating that these reflections are magnetic. 

The integrated intensity ratio of 100 reflection to conventional 1/2 1/2 1/2 magnetic Bragg reflection is only less than 1/500. Therefore, in order to completely eliminate the doubt of double reflection (multiple scattering), we carried out the same scans with different $E_{\rm i}=13.5$ and 30.5 meV. Since multiple scattering is very sensitive to the reflection angles, it should instantly disappear for different $E_{\rm i}$. As shown in Fig.~\ref{fig:data}(c), however, the 100 reflection was observed for the different $E_{\rm i}$ again. Thus the 100 magnetic reflection undoubtedly exists. 

We also confirmed the existence of temperature-dependent reflections at 110, 112 and 221. The 110 data is shown in Fig.~\ref{fig:data}(d) as the representative. All the new reflections are described by $\vec{Q}_{\rm I}=(1,0,0)$. 

Figure~\ref{fig:data}(e) shows the data of conventional 1/2 1/2 1/2 magnetic Bragg reflection. The linewidth of the 100 reflection is as sharp as that of the 1/2 1/2 1/2 one, meaning that the correlation length of $\vec{Q}_{\rm I}$ order is longer than the instrumental resolution limit of 130 {\AA} (full width at half maximum). Figure~\ref{fig:data}(f) shows the temperature dependence of integrated intensities of 100 and 1/2 1/2 1/2 reflections. Both the magnetic reflections disappear at $T_{\rm N}=21$ K with increasing the temperature. 

In this way, we discovered the additional $\vec{Q}_{\rm I}$ magnetic long-range order, which simultaneously appears below $T_{\rm N}$ in addition to the dominant $\vec{Q}_{\rm II}$ one. 

%

We discuss the magnetic structure described by $\vec{Q}_{\rm I}$. The present experiments cannot distinguish (1,0,0) and (0,0,1) for $\vec{Q}_{\rm I}$. In the tetragonal lattice expansion~\cite{Hoshi_2007}, however, spin interaction is expected to be more dominant in the $c$ plane than along the $c$ axis; (1,0,0) is more plausible than (0,0,1). In fact, $\vec{Q}_{\rm I}$ is equal to (0,0,1) in CoO with lattice contraction~\cite{Tomiyasu_2004}. 

\begin{figure}[htbp]
\begin{center}
\includegraphics[width=3.0in, keepaspectratio]{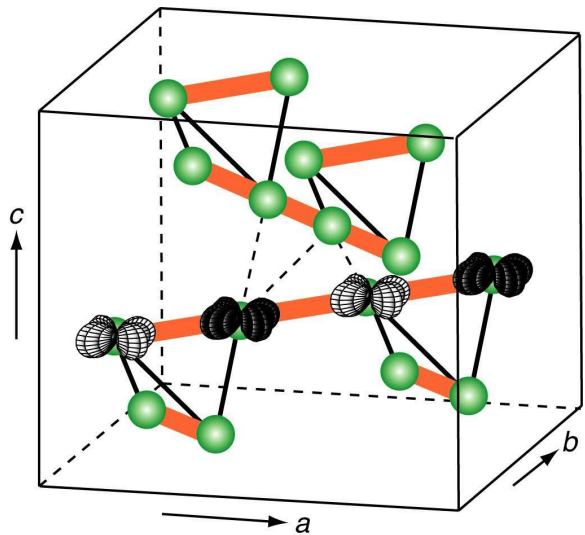}
\end{center}
\caption{\label{fig:model} (Color online) Spin and orbital model for $\vec{Q}_{\rm I}$ component in GeCo$_2$O$_4$. The orange bold lines show the antiferromagnetic spin chains connected by $d_{xy}$ orbital states. The orbitals with up and down spins (black and white polygons) on one chain are depicted as the representative. }
\end{figure}

Furthermore, recent ultrasound measurements reported that, as temperature decreases, longitudinal elastic moduli not only along the $\langle111\rangle$ direction but also the $\langle110\rangle$ direction remarkably soften towards $T_{\rm N}$~\cite{Watanabe_2008}. The softening along the $\langle111\rangle$ direction is most likely connected to the trigonal magnetic ordering by strong spin-lattice coupling~\cite{Watanabe_2008}. Although an origin of the other softening has been unresolved, this suggests that the $\vec{Q}_{\rm I}$ order is composed of spin chains along the [110] and [1$\bar{1}$0] directions in the $c$ plane, as shown in Fig.~\ref{fig:model}(a). 

Determination of more precise magnetic structure is difficult owing to the complex structure of pyrochlore lattice, the multi-domains and the neutron absorption effect of Co nuclei. Even for an isomorphic cubic material GeNi$_2$O$_4$ with no strong absorption, single-crystal polarized neutron diffraction experiments have recently determined its magnetic structure at last.~\cite{Matsuda_2008} 

The $\vec{Q}_{\rm I}$ magnetic order in GeCo$_2$O$_4$ can be understood in the standard orbital picture under tetragonal lattice expansion described as $(d_{yz})^2$, $(d_{zx})^2$, $(d_{xy})^1$, $(d_{3z^2-r^2})^1$ and $(d_{x^2-y^2})^1$. An unpaired electron occupies $d_{xy}$ without orbital degree of freedom in $t_{2g}$ states below $T_{\rm N}$. This electron will lead to kinetic direct exchange between the nearest neighbor Co$^{2+}$ ions~\cite{Watanabe_2008}, and will promote the formation of spin chains shown in Fig.~\ref{fig:model}. 
If so, however, it is interesting that a Co$^{2+}$ ion would exhibit the character of $d_{xy}$ orbital, normally vanished by spin-orbit coupling like in CoO ($S=3/2$, $L=1$, $m\sim4$ $\mu_{\rm B}$).~\cite{Kanamori_1957,Tomiyasu_2006} In fact, the Curie-Weiss fitting to magnetic susceptibility data gives us magnetic moment of about 4.9 $\mu_{\rm B}$ for GeCo$_2$O$_4$, which is much larger than the spin-only value 3 $\mu_{\rm B}$~\cite{Lashley_2008}. Therefore, it seems that the $LS$ orbital state ($L\sim1$ $\mu_{\rm B}$) with $\vec{Q}_{\rm II}$ component is dominant, and that the non-$LS$ ($d_{xy}$) character with $\vec{Q}_{\rm I}$ component supplementarily adds to the $LS$ state. 

We consider why the spin-orbit coupling tends to be inactive in GeCo$_2$O$_4$ compared to in CoO. The Co$^{2+}$ ions with $S=3/2$ have a hole in $t_{2g}$ orbital states, which generates $\vec{L}$ characters~\cite{Kanamori_1957}. Since the former and latter materials undergo tetragonal lattice {\it expansion} and {\it contraction} below $T_{\rm N}$, respectively~\cite{Hoshi_2007,Tombs_1950,Germann_1974}, the hole will occupy $d_{xy}$ in the former one and degenerated $d_{yz}$ or $d_{zx}$ in the latter one. 
Furthermore, the $d_{xy}$, $d_{yz}$, $d_{zx}$, $d_{x^2-y^2}$ and $d_{3z^2-r^2}$ states are connected to $\vec{L}$ states by the following equations, 
$d_{xy}=(|2\rangle - |{-2}\rangle)/\sqrt{2}$, 
$d_{yz}=(|1\rangle + |{-1}\rangle)/\sqrt{2}$, 
$d_{zx}=(|1\rangle - |{-1}\rangle)/(-\sqrt{2})$, 
$d_{x^{2}-y^{2}}=(|2\rangle + |{-2}\rangle)/\sqrt{2}$ 
and $d_{3z^2-r^2}=|0\rangle$, 
where the kets mean $|L_{z}\rangle$ states with $L=2$~\cite{Kanamori_1957}. Therefore, unless $d_{x^{2}-y^{2}}$ is degenerated with $d_{xy}$, the spin-orbit coupling cannot split $d_{xy}$ to the pure $\vec{L}$ states of $|2\rangle$ and $|{-2}\rangle$, which is consistent with the inactivity. In contrast, the degenerated $d_{yz}$ and $d_{zx}$ states can transform to $|1\rangle$ and $|{-1}\rangle$ by linear combination between themselves. 

%
In summary, by single-crystal neutron diffraction on GeCo$_2$O$_4$, we discovered an additional $\vec{Q}_{\rm I}$ magnetic long-range order, which simultaneously emerges at $T_{\rm N}$ with the dominant $\vec{Q}_{\rm II}$ one. The $\vec{Q}_{\rm I}$ vector matches the tetragonal lattice deformation below $T_{\rm N}$ in symmetry. This discovery suggests the appearance of $t_{2g}$ character in Co$^{2+}$, normally vanished by spin-orbit coupling like in CoO.

\begin{acknowledgments}
We thank Mr. M. Ohkawara, Mr. K. Nemoto, Mr. M. Onodera and Mr. T. Asami for their supports at JAEA and Tohoku University. The neutron experiments were performed under User Programs conducted by ISSP, University of Tokyo. This work was supported by the MEXT of Japan, Grants in Aid for Young Scientists (B) (22740209), Priority Areas (22014001), Scientific Researches (S) (21224008) and (A) (22244039) and Innovative Areas (20102005), and by Tohoku University, Inter-university Cooperative Research Program of the Institute for Materials Research. 
\end{acknowledgments}

\bibliography{GeCo2O4_el_arXiv}

\end{document}